\documentclass[a4paper,10pt,twoside]{cpc-hepnp}

\usepackage{multicol}
\usepackage{graphicx}
\usepackage{booktabs}
\usepackage{amssymb,bm,mathrsfs,bbm,amscd}
\usepackage[tbtags]{amsmath}
\usepackage{lastpage}

\begin{document}

\fancyhead[c]{\small submitted to 'Chinese Physics C'
} \fancyfoot[C]{\small 010201-\thepage}

\title{The simulation of loss of U ions due to charge changing processes in the CSRm ring\thanks{Supported by National Natural Science
Foundation of China (11305227) }}

\author{%
      ZHENG Wen-Heng$^{2;1;1)}$\email{zhengwh@impcas.ac.cn}%
\quad YANG Jian-Cheng$^{1}$
\quad LI Peng$^{1}$
\and LI Zhong-Shan$^{2}$
\quad SHANG Peng$^{2}$
\quad QU Guo-Feng$^{2}$
\and GE Wen-Wen$^{2}$
\quad TANG Mei-Tang$^{2}$
\quad SHA Xiao-Ping$^{1}$
}
\maketitle

\address{%
$^1$ Institute of Modern Physics, Chinese Academy of Sciences, Lanzhou 730000, People＊s Republic of China\\
$^2$ University of Chinese Academy of Sciences, Beijing 100049, People＊s Republic of China\\
}

\begin{abstract}
Significant beam loss caused by the charge exchange  processes and ions impact induced outgassing play a crucial role in the limitation of the maximum number of accumulated heavy ions during the  high intensity operation in the accelerators. With the aim to control beam loss due to charge exchange  processes and to confine the generated desorption gas, the tracking of the loss positions and installing the absorber blocks with low-desorption rate material at appropriate locations in the CSRm ring will be taken. The loss simulation of U ions lost an electron will be presented in this report and the calculation of the collimation efficiency of the CSRm ring will be continued in the future.
\end{abstract}

\begin{keyword}
beam loss, charge exchange, CSRm, collimation
\end{keyword}

\begin{pacs}
29.20.db 41.85.si
\end{pacs}

\footnotetext[0]{\hspace*{-3mm}\raisebox{0.3ex}{$\scriptstyle\copyright$}2013
Chinese Physical Society and the Institute of High Energy Physics
of the Chinese Academy of Sciences and the Institute
of Modern Physics of the Chinese Academy of Sciences and IOP Publishing Ltd}%

\begin{multicols}{2}

\section{Introduction}

CSR is an essential part of the HIRFL-CSR project at IMP (Institute of Modern Physics) in China Lanzhou. It consists of a main cooler storage  ring (CSRm) and an experimental ring (CSRe). The two cyclotrons Sector Focus Cyclotron (SFC) (K=69) and Separated Sector Cyclotron (SSC) (K=450) of the HIRFL (Heavy Ion Research Facility in Lanzhou) are used as its injector system. The CSRm ring, which connects  the previous HIRFL machine with the latter experimental ring (CSRe), is designed to accumulate, electron cool  and accelerate heavy ions. The cooled beam is then trapped adiabatically into 2 RF buckets and accelerated to a middle energy platform after which the cooled beam is re-trapped adiabatically into 1 RF buckets and accelerated to the extraction energy\cite{lab1}.The layout of the CSRm ring can be seen in Fig.~\ref{fig1}.
\begin{center}
\includegraphics[width=8cm]{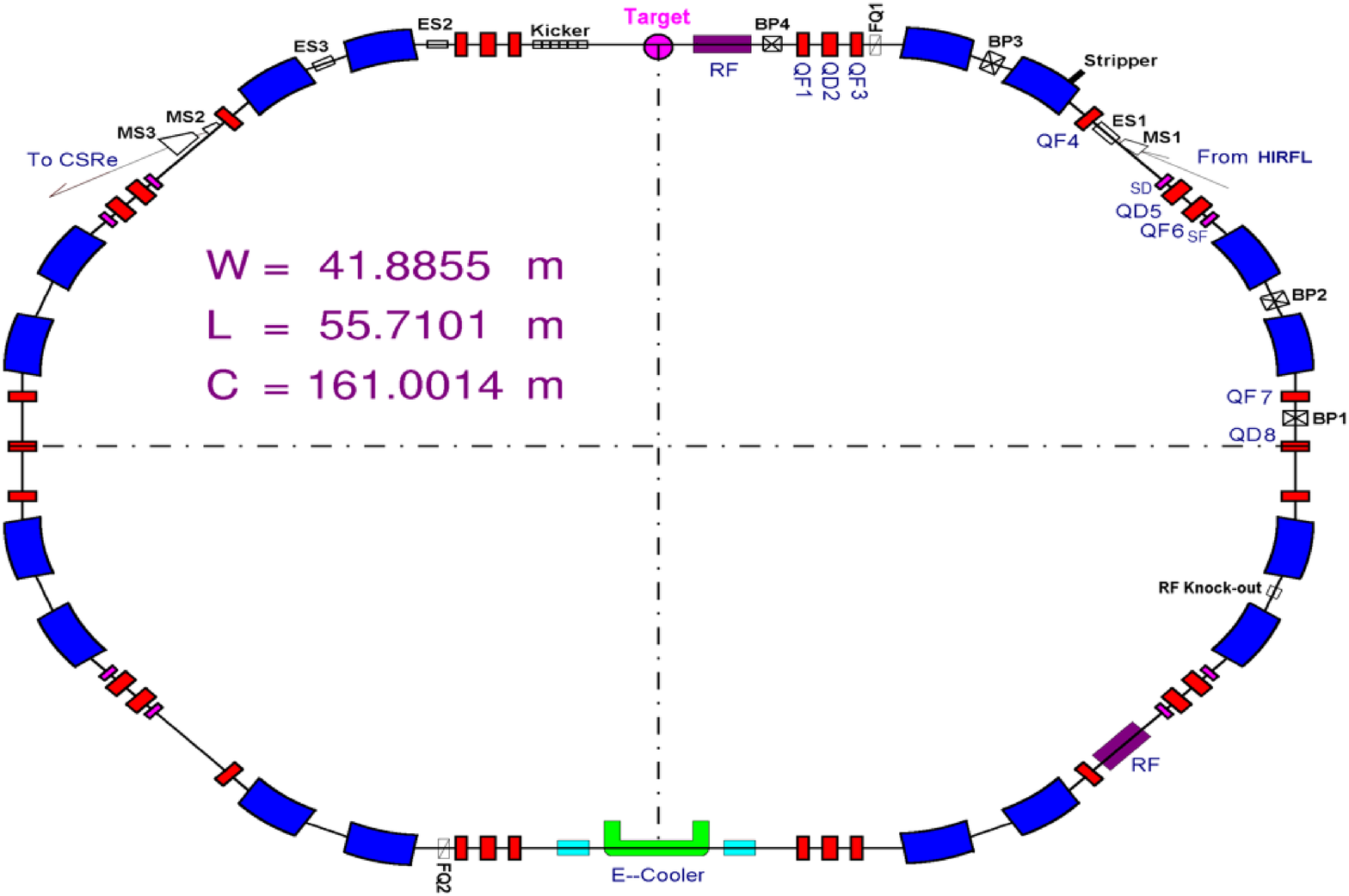}
\figcaption{\label{fig1}   The CSRm ring layout. }
\end{center}
$^{238}$U$^{32+}$ ions with the  energy of 1.237MeV/u that were  accumulated by multi-turn injection and electron cooling are chosen as the reference ion  in this paper. Plenty of life-time during the  accumulation, bunching and acceleration should be needed.

The interaction between the ions and the residual gas, which mainly consists of H$_{2}$, CO, CO$_{2}$ and CH$_{4}$ with the 10$^{-11}$ mbar ultra-high vacuum (UHV), can lead to a change of the charge state of the projectiles in CSRm ring\cite{lab2}. In the presence of dispersive ion optical elements, the trajectories of up-charged or down-charged particles are not matching one of the reference charged state, resulting in a change of the magnetic rigidity of the particle, in consequence, hit the vacuum chamber wall under grazing incidence and gas is released\cite{lab3}.Furthermore, this ion-induced desorption leads to a fast degradation of the UHV pressure during high intensity operation, which can even end up in an avalanche process and a dramatic decrease in beam life-time\cite{lab4,lab5}.

Due to the charge change-generated beam losses and the associated produced desorption gases, a simulative study of beam loss due to charge changing processes to compute and track the loss profile is indispensable in the  CSRm in this paper.   In section 2, the principle of the simulation is reported. In section 3, the simulation of loss process are presented and discussed. In section 4, the conclusion is given, a further approach to measure the desorption rate and the dynamic vacuum will be mentioned in prepare for the project of the collimation  system in the  CSRm in the future.

\section{Principle of the simulations}

Allowed to concentrate on the charge changed particles,the capture projectile ionization cross sections are much larger than the electron loss\cite{lab6}. Since the absorb blocks will be installed in both side of the horizontal orientation in this work, the difference of the projectile ionization cross sections of capture or loss electrons between the heavy ions and the residual gas is ignored. For example, the tracking of one electron-loss  $^{238}$U$^{33+}$ with the reference charge state of $^{238}$U$^{32+}$ will be focused in this paper.

The dynamics of $^{238}$U$^{33+}$ is simulated in the CSRm with the beta functions shown in Fig.~\ref{fig2}.
\begin{center}
\includegraphics[width=8cm]{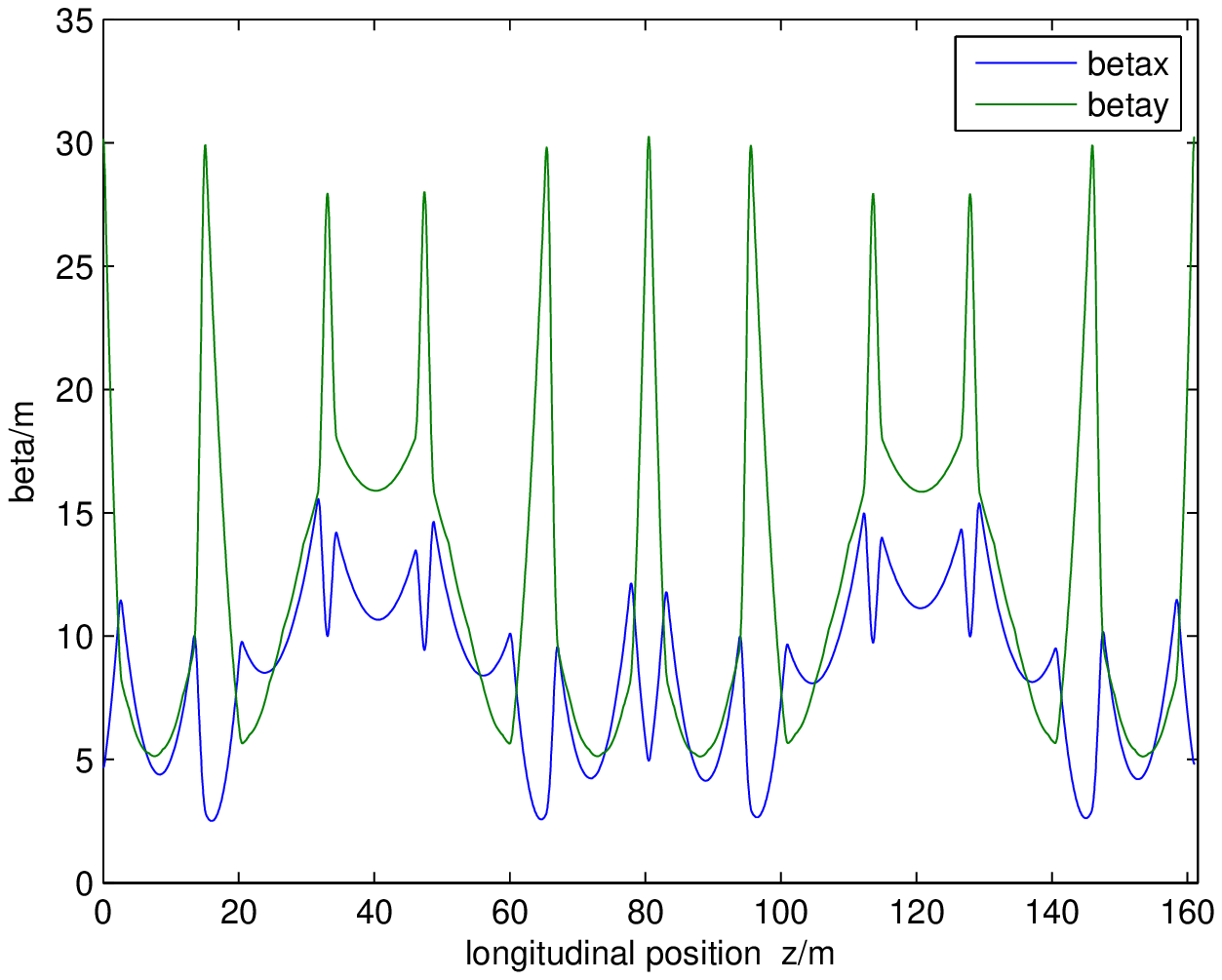}
\figcaption{\label{fig2}   Beta functions in both horizontal and vertical plan of the CSRm ring. }
\end{center}
and dispersion function in Fig. 3,respectively.
Fig.~\ref{fig3}.
\begin{center}
\includegraphics[width=8cm]{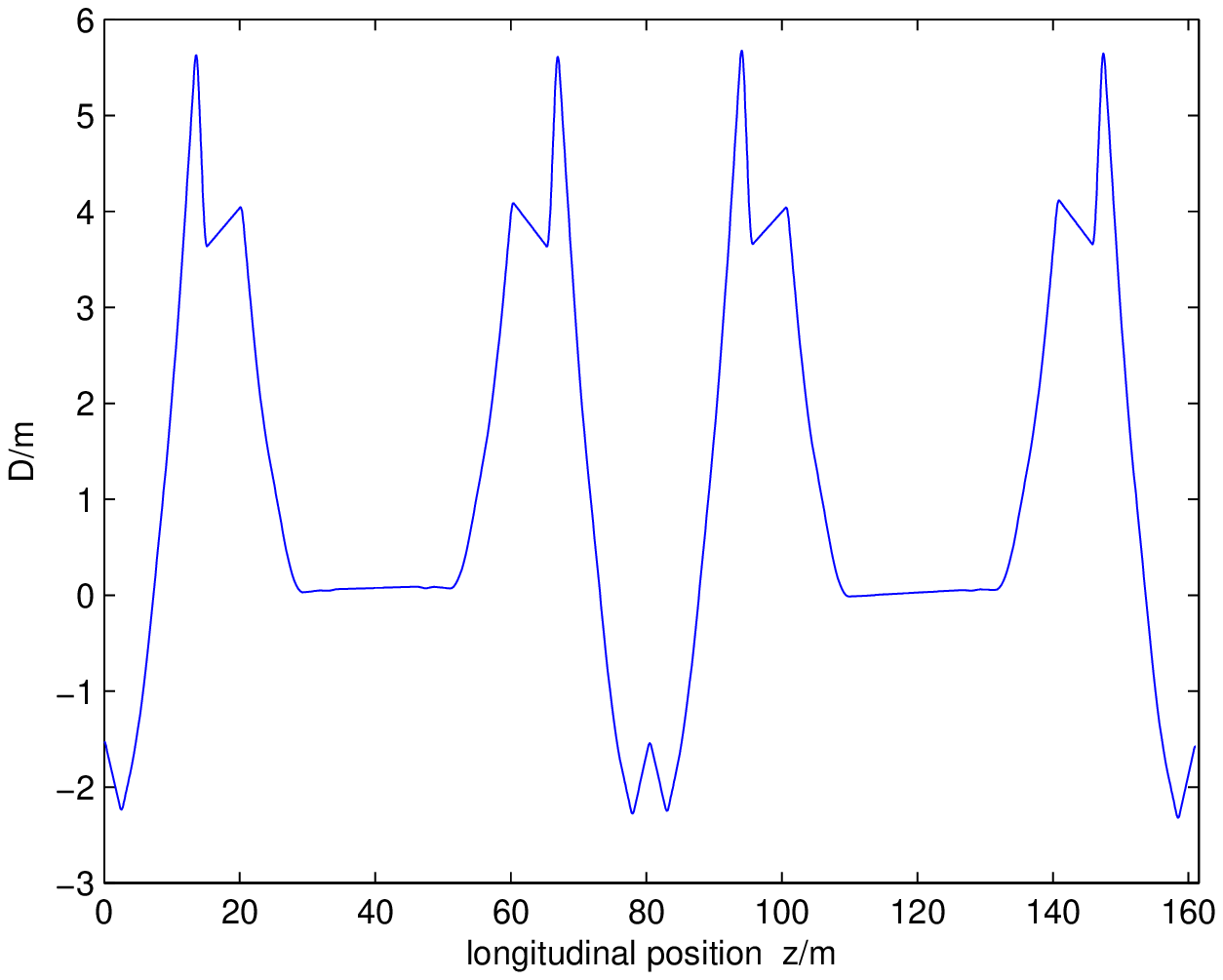}
\figcaption{\label{fig3}   Dispersion function in x plane around the CSRm ring. }
\end{center}
The beta functions and dispersion can be calculated  by the three-dimensional matrix in the transverse dimension\cite{lab7}. As $^{238}$U$^{33+}$ may be produced at any location of the element in the ring, every element of the whole lattice is divided into 10 parts uniformly with one $^{238}$U$^{33+}$ particle generated at each split point. Therefore 1200 $^{238}$U$^{33+}$ ions are generated at 1200 discrete locations along the whole ring. The initial generated $^{238}$U$^{33+}$ particles get random transverse coordinates and momentum offset on accurate matching to the lattice parameters in the 3$\sigma$ Gaussian distribution at the generated points\cite{lab8}. Specific parameters of the reference ions are listed in Table~\ref{tab1}.
\begin{center}
\tabcaption{ \label{tab1}  Beam parameter used to generate $^{238}$U$^{33+}$.}
\footnotesize
\begin{tabular*}{80mm}{c@{\extracolsep{\fill}}ccc}
\toprule Parameter & Value \\
\hline
Circumference/m & 161.0014 \\
Energy of referenced U32+ ions/(MeV/u) & 1.237 \\
Horizontal physical admittance4考/(羽mmmrad) & 150 \\
Horizontal physical emittance3考/(羽mmmrad) & 84\\
vertical physical admittance4考/(羽mmmrad) & 75\\
Vacuum/mbar & 3.0℅10$^-$$^1$$^1$\\
Momentum spread & 0.2℅10$^-$$^3$\\
Momentum offset & 3.1℅10$^-$$^3$\\
Tune values$(Q_x/Q_y)$ & 3.695/2.73\\
Dipole/mm$^2$ & 140x60\\
Quadruple/mm$^2$ & 160x100\\
Drift/mm$^2$ & 160x100\\
\bottomrule
\end{tabular*}
\vspace{0mm}
\end{center}
The Gaussian distribution can be formed by Box-Muller method. The initial randomly produced 10000 numbers $\zeta_1$ $\zeta_1$ between 0 and 1 as the uniform distribution and the Gaussian distribution $\zeta$ can be written as
\begin{eqnarray}
\label{eq1}
\bm\zeta={\sqrt{-2\ln {\bm\zeta_{1}}}{\sin{2\pi \bm\zeta_{2}}}}.
\end{eqnarray}
The Gaussian distribution matched the lattice parameters can be written as
\begin{eqnarray}
\label{eq2}
x_{\bm\beta}={\sqrt{\frac{\bm\epsilon}{\bm\gamma}}  \bm\zeta_{1}}-\frac{\bm\alpha}{\bm\gamma}{\sqrt{\frac{\bm\epsilon}{\bm\beta}} \bm\zeta_{2}}.
\end{eqnarray}
\begin{eqnarray}
\label{eq3}
x'_{\bm\beta}={\sqrt{\frac{\bm\epsilon}{\bm\beta}} \bm\zeta_{2}}.
\end{eqnarray}
where x$_\beta$ x'$_\beta$ represent the particle's horizontal position and radian between the horizontal and longitudinal, respectively\cite{lab9}.The deviation of the m/q ratio of a particle with a different charge state q compared to the reference ion with the charge state q$_0$ is equivalent to a momentum deviation 忖p/p of
\begin{eqnarray}
\label{eq4}
{\frac{\bm\Delta p}{p}}={\frac{q_{0}}{q}-1}.
\end{eqnarray}

On this occasion the behavior of $^{238}$U$^{33+}$ particles transferred in the lattice is simulated by adding $\Delta$p/p=(32-33)/33 to the relative momentum offset, compared to which the initial momentum offset can be neglected\cite{lab10}.

Within the scope of the horizontal available aperture, all the  particles are tracked from their respective generated points along the whole lattice separately according to the three-dimensional transfer matrix in transverse beam dynamics\cite{lab7}. Some complicated elements such as sextupole, kicker and so on  are replaced by simple horizontal ones in the simulation  and the vacuum pressure keeps constant. During the simulation, the tracking of the particles  which beyond the available rectangular aperture will be stropped and the particle number  will be counted at the lost  position. However, the simulation for  ions with large vertical emittance which may go beyond the vertical acceptance of the ring during the injection are not contained in this calculation.

The path is plotted only within the available machine apertures according to the designed reports\cite{lab11}, which are also listed in Table 1.

The paths of generated $^{238}$U$^{33+}$ ions in the first quadrant of the CSRm ring are shown in Fig .~\ref{fig4}.
\begin{center}
\includegraphics[width=8cm]{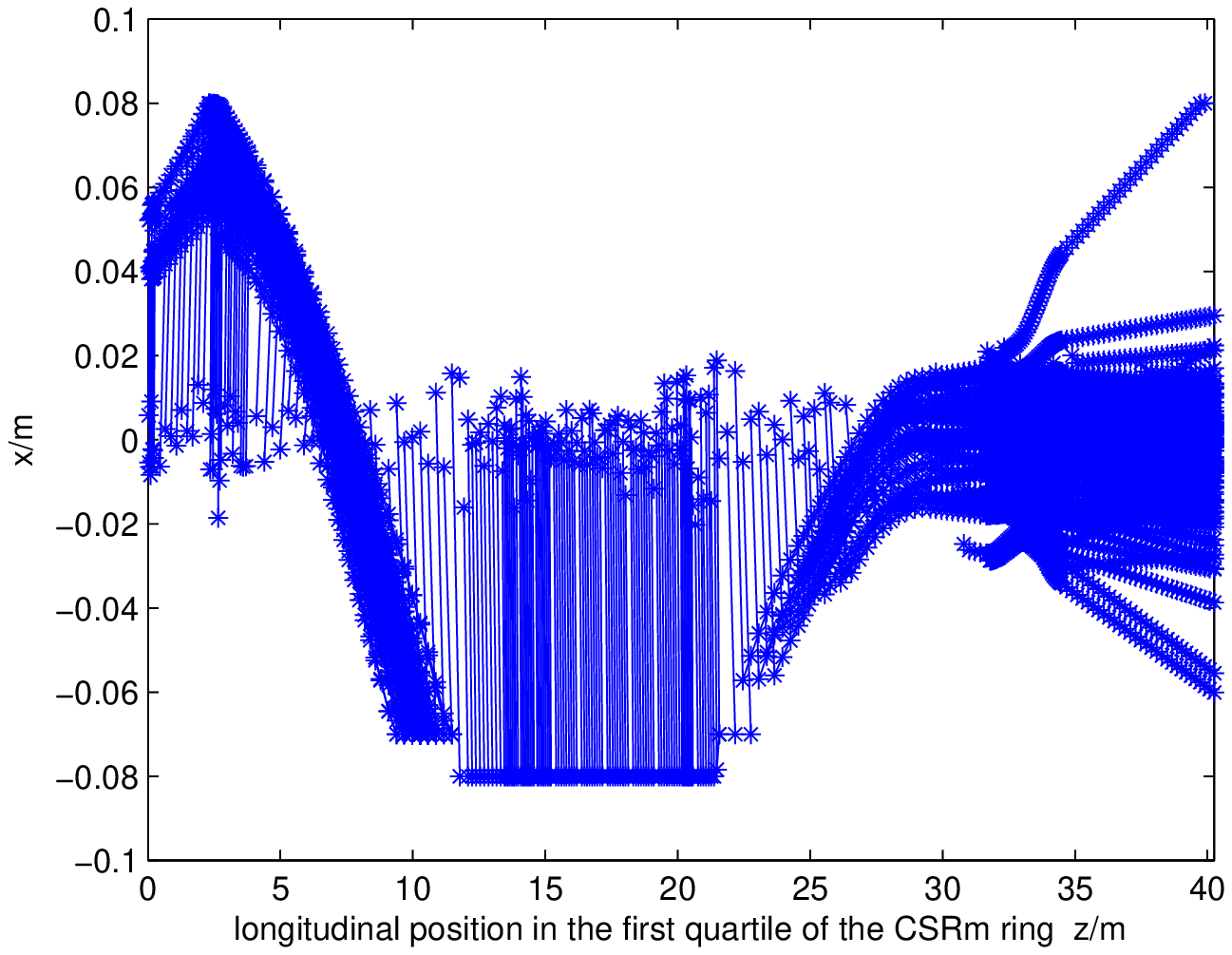}
\figcaption{\label{fig4}   $^{238}$U$^{33+}$ ions track from the point of creation until loss. }
\end{center}
The particles generated in large dispersion points will have a big offset between the initial generated point and the next transferred point, as a result of the large momentum offset.

\section{Calculation of beam loss distribution for $^{238}$U$^{33+}$ ions}

Since every ion can be tracked in the lattice, the starting point and the lost position of the lost particles can be known. The particles which beyond the available rectangular aperture in the transverse dimension  will be counted at the longitudinal position during the simulation and then  the tracking are also stopped. Thus, the lost locations with the lost particle numbers around the ring are calculated and shown in Fig.~\ref{fig5}.
\begin{center}
\includegraphics[width=8cm]{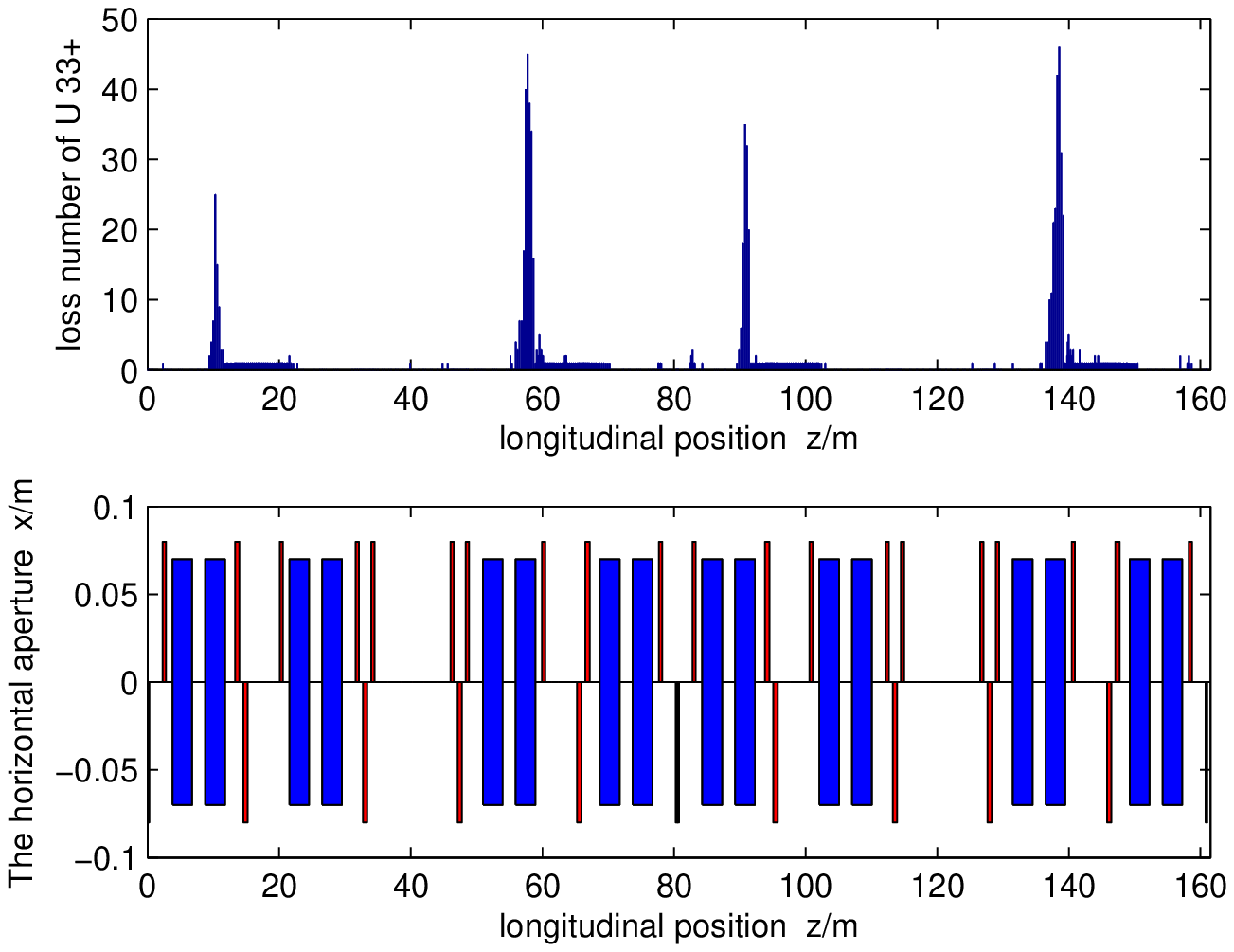}
\figcaption{\label{fig5}   Loss pattern in the CSRm lattice assuming constant pressure around the ring.}
\end{center}
during the first turn in the ring. The horizontal ordinate represents the position in CSRm, the vertical ordinate represents the lost numbers of $^{238}$U$^{33+}$ and the horizontal aperture of the dipole (blue) and quadruple (red) in Fig. 5, respectively. The focusing quadruples are plotted upper with the defocusing ones downward and the drift are replaced by the straight lines in lattice in Fig. 5.

The phenomena and results observed from the figure are summarized as follow:

1)Most charge changing  particles are lost  during their first loop in the ring.

2)Most charge changing  particles likely are lost   in the second dipole of the four super-period structures.

3)The changing charge state particles, generated in the straight drift section with a large dispersion between the second and third dipole in every super-period structure, will mostly all lose in the straight section before it reaches the adjacent bending magnet.

4)The changing charge state particles generated in the dispersion free sections are lost in the second dipole in the following super-period structure.

5)The changing charge state particles generated in the second group of the latter dipoles mostly are lost in the first group of the following super-period structure, with a few lost in the dispersion free straight section or the focusing quadruples right behind it.

6)Once the loss profile of the particles can be calculated, locations of low outgassing absorber blocks are more valid and efficient by setting them at the place of large number loss particles, the collimation efficiency can also be improved\cite{lab7}.

\section{Conclusion and future plan}

The simulation of loss distribution of one electron-loss $^{238}$U$^{33+}$ ions with the reference $^{238}$U$^{32+}$ ions has been calculated around CSRm. The correspondingly and randomly generated particles along the longitudinal locations are tracked. And the conclusion is that the lost particles are most found in the second dipole of the first group of the two adjacent dipoles in the super-period structure. The loss profile of $^{238}$U$^{33+}$ ions provides a possible way for the effective locations of low outgassing absorber blocks and also lays a foundation for the prospective collimator system in CSRm.

The work will be continued with the probable installation of low outgassing absorber blocks in the following calculation . The calculation of collimation  efficiency can be implemented  by obtain the ratio of the number of particles hitting the collimators (low outgassing absorber blocks) divided by the total number of lost particles\cite{lab2}. The work will be further worked on the desorption measurements with the prototype collimator  being designed and the simulation included the dynamic vacuum as well in CSRm in following project.

\vspace{-1mm}
\centerline{\rule{80mm}{0.1pt}}

\end{multicols}

\clearpage

\end{document}